\documentclass[a4paper,10pt,twoside]{cpc-hepnp}
\usepackage{CJK,upgreek,fancyhdr}
\usepackage{multicol}
\usepackage{graphicx}
\usepackage{booktabs}
\usepackage{amssymb,bm,mathrsfs,bbm,amscd}
\usepackage[tbtags]{amsmath}
\usepackage{lastpage}
\usepackage{color}

\usepackage{epstopdf}
\usepackage{hyperref}
\hypersetup{colorlinks=true, linkcolor=blue, urlcolor=blue, citecolor=red, bookmarks=false, pdfstartview={FitH}}

\begin{document}
\begin{CJK*}{GB}{gbsn}

\fancyhead[c]{\small Chinese Physics C~~~Vol. xx, No. x (201x) xxxxxx}
\fancyfoot[C]{\small 010201-\thepage}

\footnotetext[0]{Received 31 June 2015}

\title{Scattering of massless fermions by Schwarzschild and Reissner-Nordstr\"om black holes}

\author{ Ciprian A. Sporea$^{1;1)}$\email{ciprian.sporea@e-uvt.ro} }

\maketitle

\address{ $^1$ West University of Timi\c soara, V.  P\^ arvan Ave.  4, RO-300223 Timi\c soara, Romania }

\begin{abstract}
In this paper we are studying the scattering of massless Dirac fermions by Schwarzschild and Reissner-Nordstr\"om black holes. This is done by applying the partial wave analysis to the scattering modes obtained after solving the massless Dirac equation in the asymptotic regions of the two black hole geometries. We succeed to obtain analytic phase shifts with the help of which the scattering cross section is computed. The glory and spiral scattering phenomena are showed to be present like in the case of massive fermion scattering by black holes.
\end{abstract}



\footnotetext[0]{\hspace*{-3mm}\raisebox{0.3ex}{$\scriptstyle\copyright$}2013
Chinese Physical Society and the Institute of High Energy Physics
of the Chinese Academy of Sciences and the Institute
of Modern Physics of the Chinese Academy of Sciences and IOP Publishing Ltd}%

\begin{multicols}{2}

\section{Introduction}

The problem of black hole scattering is still an ongoing on, despite the numerous papers that have dealt with it so far. Most of these studies have been dedicated to investigating different aspects of (massless) scalar wave scattering by black holes \cite{Collins,Ford,Zhang,Matzner,Matzner1,Matzner2,Sanchez,Higuchi,Jung2,Jung3,Macedo,crispino,Benone,Batic,Kanai, Gubmann,Macedo2,Futterman,Unruh,Anderson,Darwin, Gaina1,Gaina2}. However, there are also papers that studi the scattering of massless electromagnetic waves \cite{mashh,Frolov1,Fabbri,crispinoEM1,Kobayashi,Konoplya,crispinoEM2,crispinoEM3,crispinoEM4} and gravitational waves \cite{Vish,Westervelt,Peters,Handler,dolanG1, dolanG2, Kanti,Sorge}. The problem of scattering of massive spinor $1/2$ waves by black holes was discussed in \cite{Unruh,dolan, Jin, ChaoLin, Doran,Huang,Das,Jung,Rogatko, sporea1, sporea2}.The scattering of massless fermions was studied in \cite{Liao,Ghosh}, where the scattering of massless fermions by a black hole with a cosmic string, respectively a dilatonic black hole was investigated   and partially in \cite{dolan} for a Schwarzschild black hole. The authors of \cite{dolan} have used numerical methods to solve the Dirac equation in Schwarzschild black hole geometries in order to find numerical phase shifts using a partial wave analysis. In \cite{sporea1} we succeeded to obtain for the first time analytic expressions for the Schwarzschild phase shifts. Furthermore, in \cite{sporea2} we extended our study to include also the case of fermion scattering by charged Reissner-Nordstr\"om black holes, where we found again analytic phase shifts. Moreover, our study \cite{sporea2} is up to our knowledge the first one in the literature in which the problem of massive fermion scattering by Reissner-Nordstr\"om black holes was investigated.

In this paper we are studying the scattering of massless fermions by spherical symmetric black holes, with a focus on Schwarzschild and charged Reissner-Nordstr\"om black holes. We derive analytic phase shifts that will allow us to write down analytic expressions for the scattering cross sections.

Although it is generally assumed that astrophysical black holes are electrically neutral \cite{Gibbons}, or at least have negligible charges, it has been recently showed in Ref. \cite{Cardoso} that charged astrophysical black holes can in fact exist in the context of minicharged dark matter models \cite{Rujula}. These dark matter models predict new fermions that can have fractional electric charge or fermions that are charged under a $U(1)$ hidden symmetry. Because these new charges have just a small fraction of the electron's charge, their coupling with the Standard Model electromagnetic sector is suppressed. This means that even in the case of massive fermions there could be no direct interaction between the Dirac field and the electromagnetic field of the black hole. This could open new possibilities for dark matter detection through neutrino-wave scattering by black holes, besides the gravitational-wave signatures discussed in \cite{Cardoso1}.

In the original Standard Model of particle physics neutrinos are assumed to be massless. Thus our results will contain also the case of massless neutrino scattering by black holes (see also Appendix A). However, it has been observed experimentally \cite{neutrino1,neutrino2} that neutrinos have a nonzero mass that it is currently bound to $\sum m_\nu < 0.183$ eV \cite{neutrinomass}. The electron neutrino mass could be as small as $m_{\nu_e}\sim0.01$ eV or smaller. Having this in mind one can easily assume that our results (presented in the following sections) will give also a good approximation for the cross sections in the case of scattering of astrophysical neutrinos by black holes.

The paper is organised as follow: in Sec. \ref{section2} we give a brief introduction of the general form of the massless Dirac equation in a curved spacetime with spherical symmetry. The Cartesian gauge is introduced and the separation of spherical variables is made. In the next Sec. \ref{section3} we solve the massless Dirac equation in the case of Schwarzschild and Reissner-Nordstr\"om black hole geometries. We focus only on scattering mode solutions. The following Sec. \ref{section4} deals with the partial wave analysis, where we give analytical expressions for the phase shifts that enter into the definitions of the scattering amplitudes and cross sections. Sec. \ref{section5} is dedicated to the presentations of the main results obtained. The paper ends with Sec. \ref{section6} where the final conclusions and some remarks are presented.

In this paper we set $G=c=\hbar=1$; the metric signature used is $(+,-,-,-)$; the natural indices are labeled with Greek letters $\mu, \nu, \alpha, ...$, while the local indices are labeled with $a, b, c, ...$; both can take values in the range $(0,1,2,3)$.

\section{Preliminaries}\label{section2}

Starting from the gauge invariant action
\begin{equation}\label{ec1}
\mathcal{S} =\int \! \mathrm{d^4}x\ \! \sqrt{-g}\left\{ \frac{i}{2}\overline\psi\gamma^a D_a\psi - \frac{i}{2}(\,\overline{D_a\psi}\,)\gamma^a\psi \right\}
\end{equation}
we can immediately derive the following Dirac equation for a massless spinor field in a curved spacetime
\begin{equation}\label{ec2}
i\gamma^a e^\mu_a\partial_\mu\psi + \frac{i}{2}\frac{1}{\sqrt{-g}}\partial_\mu(\sqrt{-g}\,e^\mu_a)\gamma^a\psi -\frac{1}{4}\{\gamma^a , S^{\,b}_{\,\,\,c} \}\omega^{\,c}_{\, a b}\psi = 0
\end{equation}
where $e^\mu_a$ are the tetrad fields such that $g^{\mu\nu}=\eta^{a b}e_a^\mu e_b^\nu$; $\gamma^a$ are the point-independent Dirac matrices satisfying $\{\gamma^a,\gamma^b\}=2\eta^{a b}$; $S^{\,b}_{\,\,c}$ are the generators of the spinor representation of $SL(2,\mathbb{C})$ \cite{Thaller} such that $S^{a b}=\frac{i}{4}[\gamma^a,\gamma^b]$;

The covariant derivative $D_a$ and the spin-connection read
\begin{equation}\label{ec3}
\begin{split}
& D_a = \partial_a + \frac{i}{2}S^{\,b}_{\,\,c}\,\omega^{\,c}_{\, a b}\\
& \omega^{\,c}_{\, ab} = e^\mu_a e^\nu_b \left( \hat e^c_\lambda\Gamma^\lambda_{\mu\nu} - \hat e^c_{\nu,\mu} \right)
\end{split}
\end{equation}
with $\partial_a=e^\mu_a\partial_\mu$ and $\Gamma^\lambda_{\mu\nu}$ stand for the GR Christoffel symbols.

The spacetime geometry of a spherically symmetric black hole is given by the following line element
\begin{equation}\label{ec4}
ds^2=h(r) dt^2-\frac{dr^2}{h(r)}-r^2\left( d\theta^2+\sin^2\theta d\phi^2 \right)
\end{equation}

In the Cartesian gauge \cite{cota1, cota2}, the above line element can be obtained from $ds^2=\eta_{ab}\hat e^a\hat e^b$ with the following choice of the tetrad field $\hat e^a(x)=\hat e^a_\mu dx^\mu$ (i.e. the 1-forms)
\begin{equation}\label{ec5}
\begin{split}
&\hat e^0 = h(r)dt \\
&\hat e^1 = \frac{1}{h(r)}\sin\theta\cos\phi\,dr + r\cos\theta\cos\phi\,d\theta - r\sin\theta\sin\phi\,d\phi\\
&\hat e^2 = \frac{1}{h(r)}\sin\theta\sin\phi\,dr + r\cos\theta\sin\phi\,d\theta + r\sin\theta\cos\phi\,d\phi\\
&\hat e^3 =\frac{1}{h(r)}\cos\theta\,dr - r\sin\theta\,d\theta
\end{split}
\end{equation}
The main advantage of the above Cartesian gauge consists in the fact that in this gauge the Dirac equation (free or in central scalar potentials) is manifestly covariant under rotations. This means that the angular part of the equation can be solved in terms of the usual 4-component angular spinors form special relativity $\Phi^\pm_{m,\kappa}(\theta,\phi)$ \cite{Thaller,Landau}. Using the Cartesian gauge new (exact or approximative) analytical solutions of the Dirac equation in curved backgrounds were found \cite{cota,cota3,sporea3,cota4}.

Inserting the metric (\ref{ec4}) in eq. (\ref{ec2}) and after some calculations, one can put the Dirac equation into a Hamiltonian form $H\psi(x)=i\,\partial_t\,\psi(x)$, with
\begin{equation}\label{ec6}
H_D=-i\gamma^0\left(\vec\gamma\cdot\vec e_r\right)\left(h\partial_r +\frac{h}{r} -\frac{\sqrt{h}}{r}K \right)
\end{equation}
and where $K=2\vec S\cdot\vec L +1 = J^2 - L^2 + \frac{1}{4}$ is the spin-orbit operator \cite{Thaller}, whose eigenvalues $\kappa$ are related to the ones of the total angular momentum operator ($J$) and of the orbital angular momentum ($L$) by
\begin{equation}\label{ec7}
\kappa=\left\{\begin{array}{lcc}
-(j+\frac{1}{2})=-(l+1)&{\rm for}& j=l+\frac{1}{2}\\
+(j+\frac{1}{2})=l&{\rm for}& j=l-\frac{1}{2}
\end{array}\right.
\end{equation}

The radial part of the Dirac equation can be derived by searching for (particle-like) positive frequency solutions of energy $E$ of the type
\begin{equation}\label{ec8}
\begin{split}
\psi (x)=&\psi_{E,\kappa,m_{j}}(t,r,\theta,\phi)= \\
& =\frac{e^{-iEt}}{r}\left\{f^{+}_{E,\kappa}(r)\Phi^{+}_{m_{j},\kappa}(\theta,\phi)
+f^{-}_{E,\kappa}(r)\Phi^{-}_{m_{j},\kappa}(\theta,\phi) \right\}
\end{split}
\end{equation}
One can show that the final form of the equations satisfied by the radial wave-functions $f^\pm(r)$ (where for simplicity we have dropped the indices $E$ and $\kappa$) read
\end{multicols}
\begin{equation}\label{ec9}
\left(\begin{array}{cc}
0 & -h(r)\frac{\textstyle d}{\textstyle dr}+\frac{\textstyle \kappa}{\textstyle r}\sqrt{h(r)}\\
&\\
h(r)\frac{\textstyle d}{\textstyle dr}+\frac{\textstyle \kappa}{\textstyle r}\sqrt{h(r)}& 0
\end{array}\right) \left(\begin{array}{c}
f^+(r) \\
\ \ \\
f^-(r)  \end{array} \right)= E \left(\begin{array}{c}
f^+(r) \\
\ \ \\
f^-(r)  \end{array} \right)
\end{equation}
\begin{multicols}{2}

\vspace{3mm}

\ \ \\

\section{The massless Dirac equation in Schwarzschild and Reissner-Nordstr\"om geometry}\label{section3}

As already mentioned in the Introduction we are studying here only the scattering of massless fermions by black holes. For that we need first to find the scattering modes of the Dirac equation (\ref{ec9}) on which to apply the partial wave analysis (PWA) method. That will allow us to find the phase shifts and then to calculate all the physical quantities that are characteristic to the scattering phenomena. For PWA one only needs to know the asymptotic behaviour of the scattering modes. As showed bellow in the asymptotic region of both the Schwarzschild and RN back hole the Dirac equation (\ref{ec9}) can be brought to a simpler form that will allow us to solve it analytically. Furthermore, having analytical solution at our disposal will allow us to find analytical phase shifts.

In the case of a Reissner-Nordstr\"om black hole the function $h(r)$ entering the line element (\ref{ec4}) is defined by
\begin{equation}\label{d1}
h(r)=1-\frac{2M}{r}+\frac{Q^2}{r^2}=\left(1-\frac{r_+}{r}\right)\left(1-\frac{r_-}{r}\right)
\end{equation}
where $M$ is the mass of the black hole and $Q$ the electric charge. The Cauchy ($r_-$) and black hole horizon ($r_+$) radii are easily found to be  $r_{\pm}=M\pm\sqrt{M^2-Q^2}$ (provided $Q<M$). If we make $Q=0$ in (\ref{d1}) we obtain the Schwarzschild line element with $r_-=r_+=r_0=2M$.

It proves useful to introduce a convenient Novikov-like dimensionless coordinate \cite{novikov,MTW}
\begin{equation}\label{d2}
x=\sqrt{\frac{r}{r_{+}}-1}\,\in\,(0,\infty)
\end{equation}

Then the Dirac equation (\ref{ec9}) in the asymptotic region of the black hole becomes
\begin{equation}\label{ec10}
\left(\begin{array}{cc}
 \frac{\textstyle 1}{\textstyle 2} \frac{\textstyle d}{\textstyle dx} +\frac{\textstyle\kappa}{\textstyle x}
& -\varepsilon \left(x+ \frac{\textstyle 1}{\textstyle x}\right)\\
&\\
\varepsilon \left(x+ \frac{\textstyle 1}{\textstyle x}\right)
& \frac{\textstyle 1}{\textstyle 2} \frac{\textstyle d}{\textstyle dx} -\frac{\textstyle \kappa}{\textstyle x}
\end{array}\right)
\left(\begin{array}{c}
 f^+(x)\\
 \, \ \\
 f^-(x)
\end{array}\right)=0
\end{equation}
where we denoted $\varepsilon=r_{+}E$. In obtaining (\ref{ec10}) we have used a Taylor expansion with respect to $1/x$ from which we neglected the $O(1/x^2)$ terms and higher.

After putting the terms proportional with $x$ into diagonal form, using the transformation matrix
\begin{equation}\label{ec11}
M=\sqrt{\varepsilon}\left(\begin{array}{cc}
-i\,\, & \,\, i\\
1 \,\, & 1\\
\end{array}\right)
\end{equation}
that transforms $(f^+, f^-)^T\rightarrow (\hat f^+, \hat f^-)^T=M^{-1}(f^+, f^-)^T $, the final system of radial equations is obtained
\begin{equation}\label{ec12}
\begin{split}
&\frac{1}{2} \frac{d\hat f^{+}}{dx}- i \varepsilon\left(x+\frac{1}{x}\right)\hat f^{+} =\frac{\kappa}{x}\hat f^{-} \\
&\frac{1}{2} \frac{d\hat f^{-}}{dx}+ i \varepsilon\left(x+\frac{1}{x}\right)\hat f^{-} =\frac{\kappa}{x}\hat f^{+}
\end{split}
\end{equation}

The analytical solutions of the above equations can be found in terms of Whittaker $M$ and $W$ functions \cite{cota,sporea1,sporea2}
\begin{equation}\label{d3}
\begin{split}
&\hat f^+(x)=C_1\frac{1}{x}M_{\rho_+,s}(2i\varepsilon x^2) +C_2\frac{1}{x}W_{\rho_+,s}(2i\varepsilon x^2) \\
&\hat f^-(x)=C_1\frac{s-i \varepsilon}{\kappa}\frac{1}{x}M_{\rho_-,s}(2i\varepsilon x^2) - C_2\frac{1}{\kappa}\frac{1}{x}W_{\rho_-,s}(2i\varepsilon x^2)
\end{split}
\end{equation}
where the parameters $s, \rho_\pm$ are related to $\kappa$ and $\varepsilon$ by the following relations
\begin{equation}\label{d4}
s=\sqrt{\kappa^2-\varepsilon^2},\quad \rho_{\pm}=\mp\frac{1}{2}-i\varepsilon
\end{equation}

These solution are the starting point for studying the scattering phenomena by Schwarzschild and Reissner-Nordstr\"om black hole with the help of partial wave analysis. The Whittaker functions $M_{\rho_\pm,s}(2i\varepsilon x^2)=(2i\varepsilon x^2)^{s+\frac{1}{2}}[1+O(x^2)] $ are regular in $x=0$ (i.e. in $r=r_+$), while the Whittaker $W_{\rho_\pm,s}(2i\varepsilon x^2)$ are divergent as $x^{1-2s}$ if $s>\frac{1}{2}$ \cite{NST}. As showed in the Appendices of ref. \cite{sporea1} and \cite{sporea2} one must impose the asymptotic condition $C_2=0$ in order to have elastic collisions with a correct Newtonian limit for large angular momentum.

\section{Scattering cross section and phase shifts}\label{section4}

The phase shifts that result after applying the partial wave analysis \cite{sporea1,sporea2} on the scattering modes (\ref{d3}) are defined by
\begin{equation}\label{final}
S_{\kappa}=e^{2i\delta_{\kappa}}=\left(\frac{\kappa}{s-i\varepsilon}\right)\,\frac{\Gamma(1+s-i\varepsilon)}{\Gamma(1+s+i\varepsilon)} e^{i\pi(l-s)}
\end{equation}
The scattering amplitudes are defined by \cite{Landau}
\begin{eqnarray}\label{fg_partial}
f(\theta)=\sum_{l=0}^{\infty}a_l\,P_l(\cos \theta)\,, \qquad  g(\theta)=\sum_{l=1}^{\infty}b_l\,P_l^1(\cos\theta)
\end{eqnarray}
where $a_l$ and $b_l$ are the partial amplitudes
\begin{eqnarray}
a_l&=&\frac{1}{2ip}\left[(l+1)(S_{-l-1}-1)+l(S_l-1)\right] \nonumber \\
b_l&=&\frac{1}{2ip}\left(S_{-l-1}-S_l\right)
\end{eqnarray}
Putting all together one gets the differential scattering cross section
\begin{equation}\label{int}
\frac{d\sigma}{d\Omega}=|f(\theta)|^2+|g(\theta)|^2
\end{equation}

In the next section we will give a selection of our key results for the scattering of massless fermions by Schwarzschild and Reissner-Nordstr\"om black holes. In the derivation of the plots we have used a method first proposed in \cite{Yennie} and further developed in \cite{dolan,sporea1,sporea2} for improving the convergence of the partial wave series (\ref{fg_partial}).

\section{Results}\label{section5}

In Fig. \ref{fig1} we compare the scattering of massless ($v=1$) and massive ($v\neq1$) fermions by a Schwarzschild black hole ($q=0$) for a fixed value of the (frequency) parameter $ME$. The case of massive fermion scattering by Schwarzschild and Reissner-Nordstr\"om black holes was studied in more detail in our previous papers \cite{sporea1,sporea2}. Analyzing the differential cross section in the backward direction (near $\theta\approx\pi$) one can observe the presence of a minima in the scattering intensity. If the fermion is massive then the scattering intensity in the backward direction is higher compared with the massless case. Moreover, decreasing the fermion speed the minima will become eventually a maxima in the backward direction (see \cite{sporea1} for more details).

\begin{center}
\includegraphics[scale=0.45]{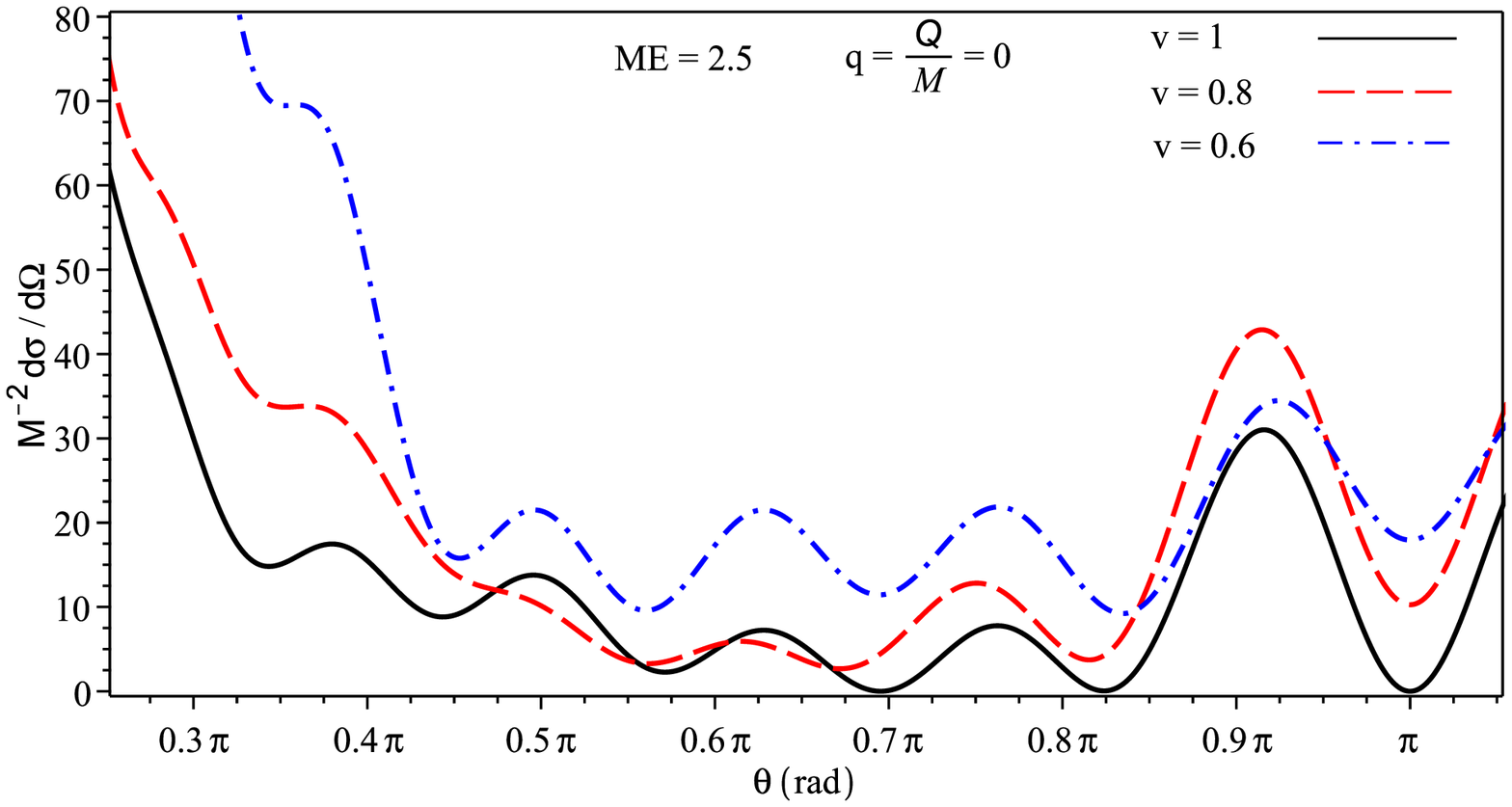}
\figcaption{\label{fig1}(color online). Comparison between the scattering of massless ($v=1$) and massive ($v\neq1$) fermions by a Schwarzschild black hole at $ME=2.5$. The presence of a minima in the backward direction can be observed. }
\end{center}

In optics the presence of a minima or maxima in the scattering intensity in the backward direction occurs when the deflection angle of a ray is a multiple of $\pi$. This is observed by the presence of a bright spot or hallo in the antipodal direction. If the orbit passes the scattering center multiple times, then spiral (or orbiting) scattering can occur. This can be seen by the presence of oscillations in the scattering intensity. As can be seen from Fig. \ref{fig1}-\ref{fig5} the phenomena of glory and spiral scattering also occurs in the case of massless fermion scattering by Schwarzschild and charged Reissner-Nordstr\"om black holes.

In Fig. \ref{fig2} we plot the scattering intensity for a massless spinor wave of fixed frequency for a Schwarzschild black hole ($q=0$), a typical Reissner-Nordstr\"om black hole (with $q=0.5,\, q=0.6,\, q=0.9$) and respectively, an extremal Reissner-Nordstr\"om black hole ($q=1$). One can observe (very clearly for $ME=3$) that at a fixed frequency the glory width gets larger as the value of the charge-to-mass ratio $q$ is increased. The same behaviour was also reported in \cite{crispino} in the case of scattering of massless scalar waves by Reissner-Nordstr\"om black holes. In Ref. \cite{Liao} the authors found that the linear mass density of the cosmic string produces a similar effect. Furthermore, the oscillations present in the scattering intensity become less frequent as we approach the extremal case $q=1$. This means that the spiral scattering becomes less important as the black hole gets more and more charge on it.

\begin{center}
\includegraphics[scale=0.45]{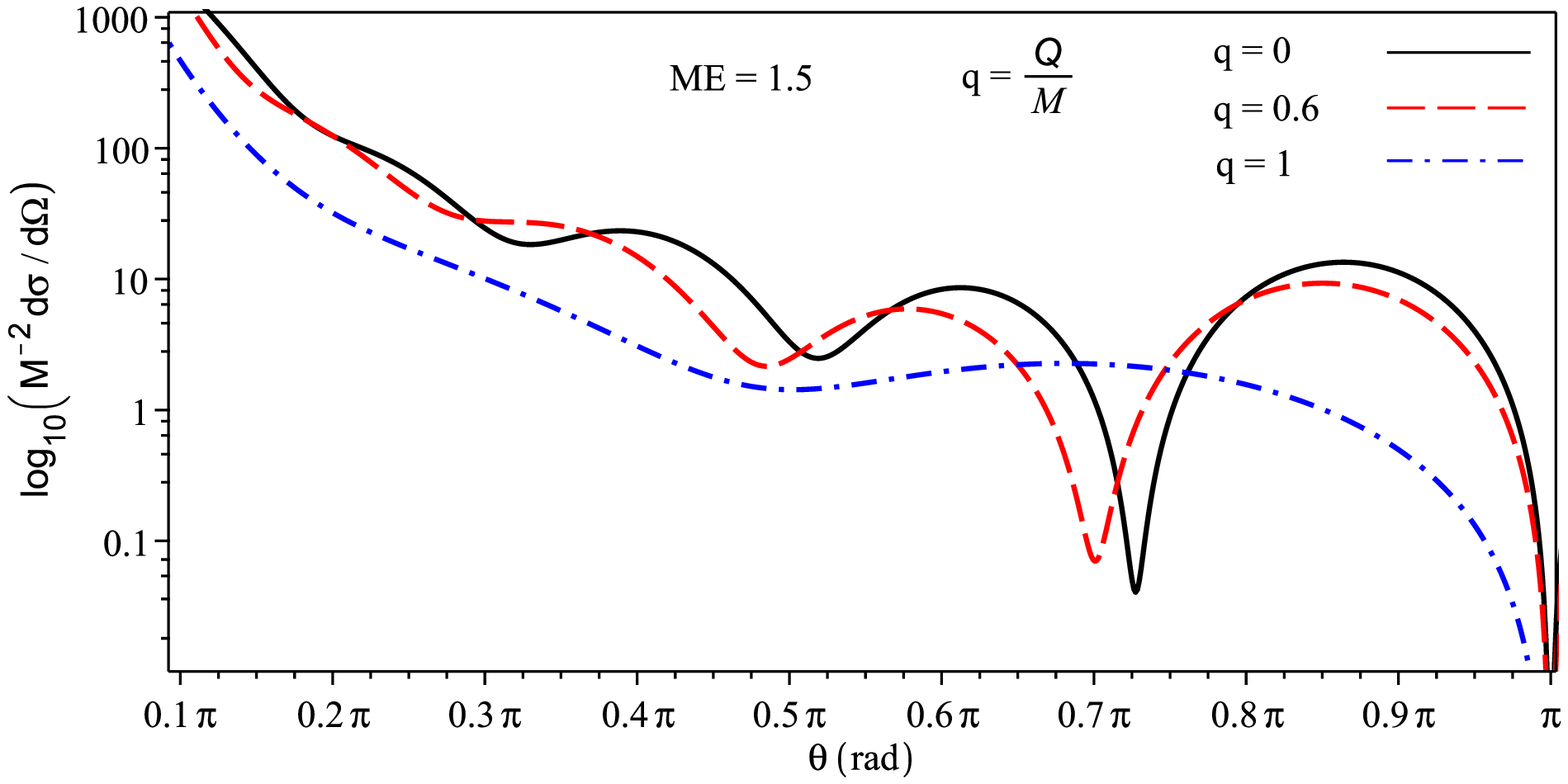}
\includegraphics[scale=0.45]{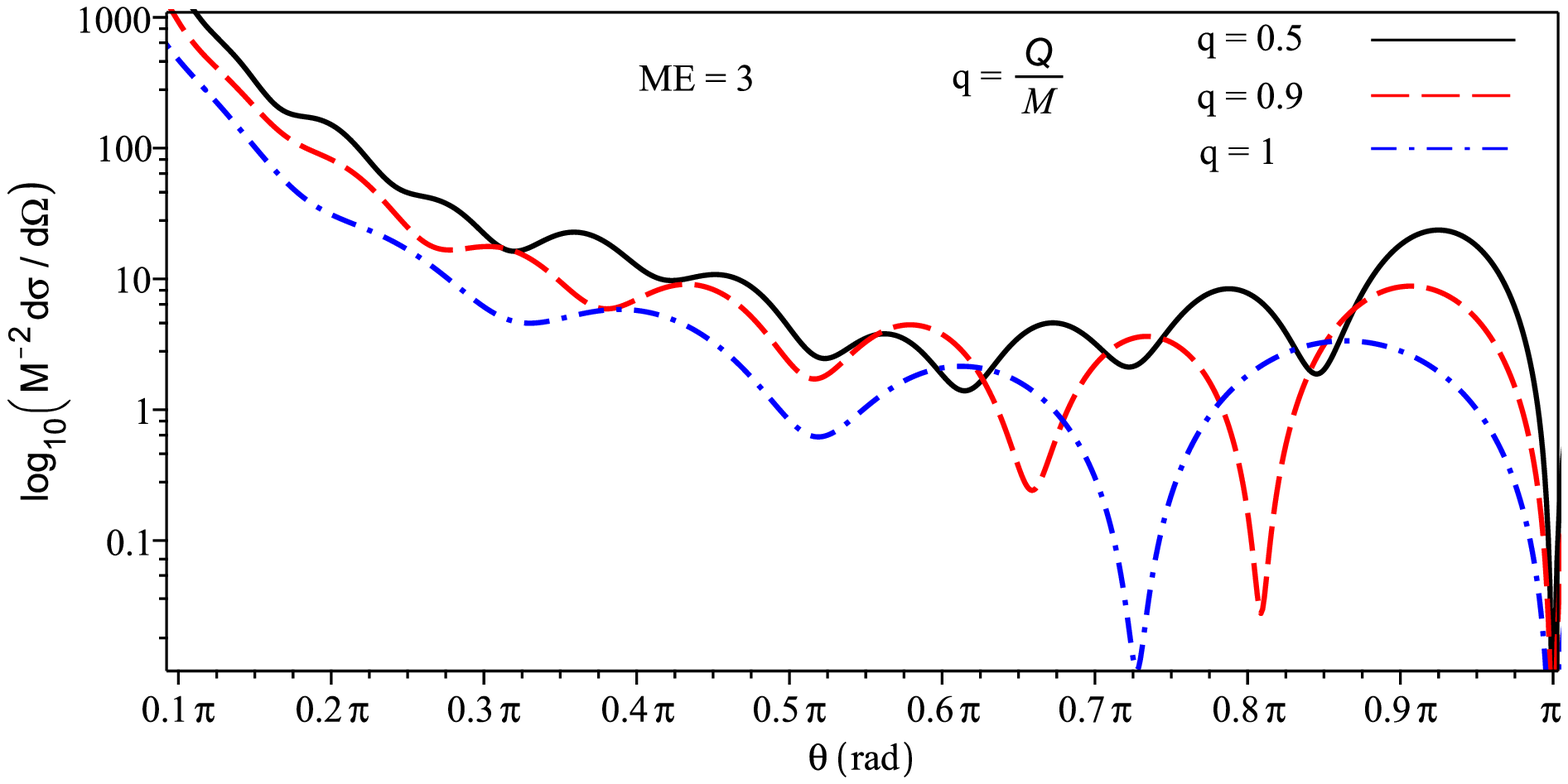}
\figcaption{\label{fig2}(color online). Comparison between the massless fermion scattering cross section at fixed frequency $ME=1.5$ for $q=0,\, q= 0.6,\ q=1$ in one case and at $ME=3$ for $q=0.5,\, q= 0.9,\ q=1$ in the other case. Fixing the frequency the glory width gets larger as the value of the charge-to-mass ratio $q$ is increased. }
\end{center}

In Fig. \ref{fig3} the differential scattering cross section for the massless fermion field is plotted, using a logarithmic scale, for different values of the incoming fermion frequency $ME=2.5,\, 3,\, 3.5$ at a chosen fixed value of the charge-to-mass ratio $q=0$ (Schwarzschild case), $q= 0.5$ (typical Reissner-Nordstr\"om) and $q=1$ (extremal Reissner-Nordstr\"om case). The first thing to observe is the fact that the width of the glory becomes narrower as the frequency increases. On the contrary the oscillations (indicating spiral scattering) present in the scattering intensity become more frequent as the value of $ME$ is increased. This can be best seen for the extremal case $q=1$.

\begin{center}
\includegraphics[scale=0.45]{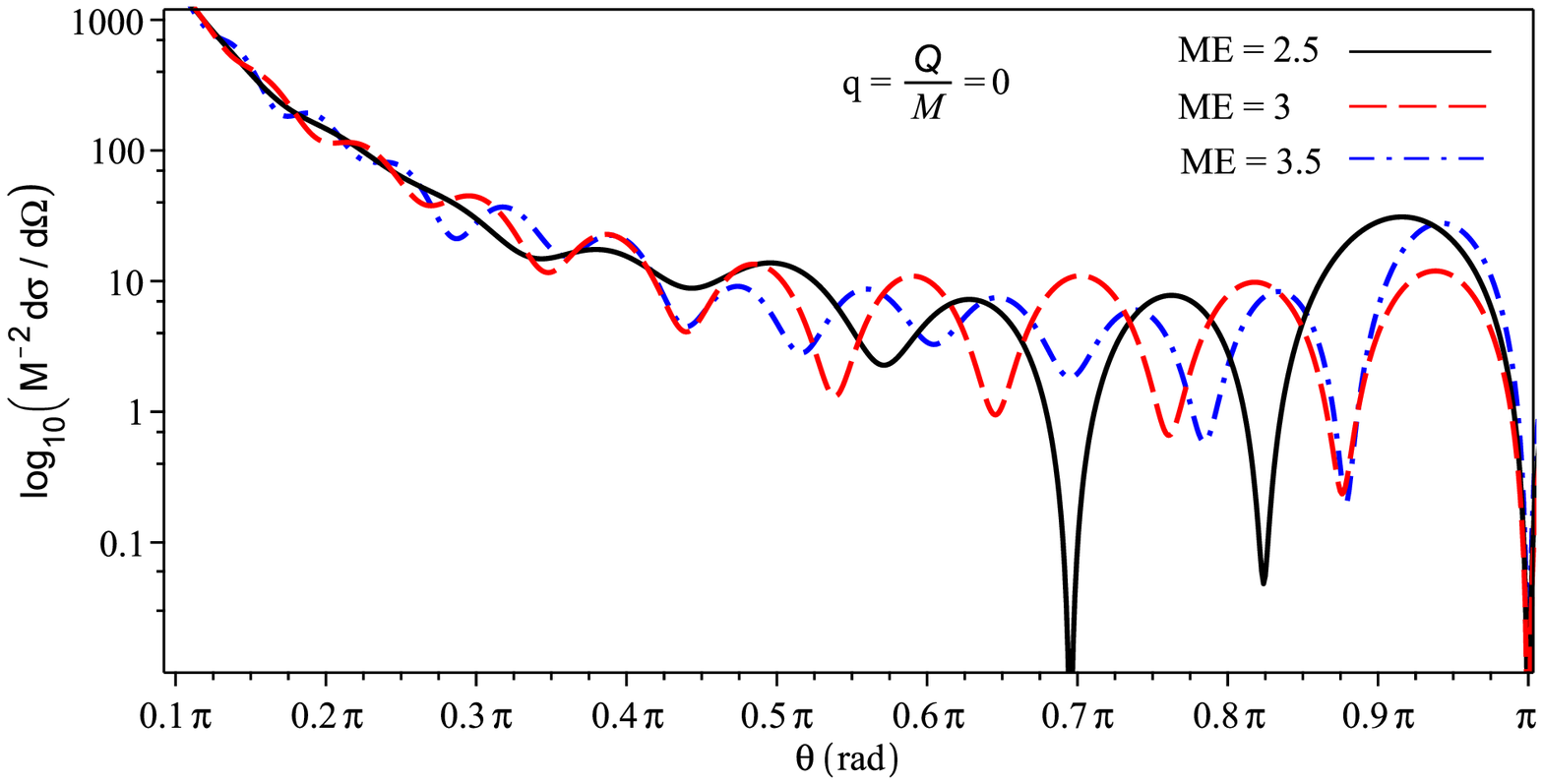}
\includegraphics[scale=0.45]{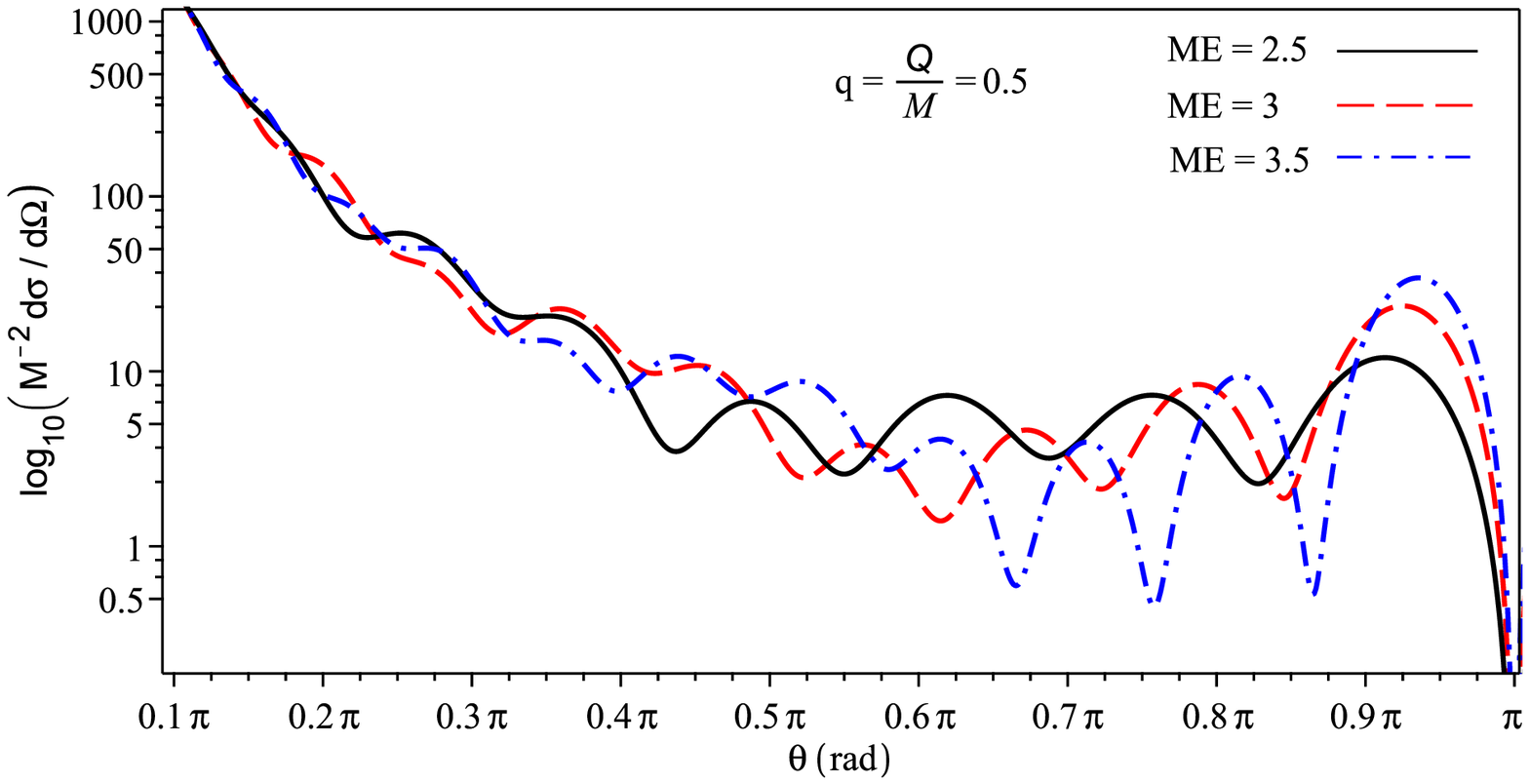}
\includegraphics[scale=0.45]{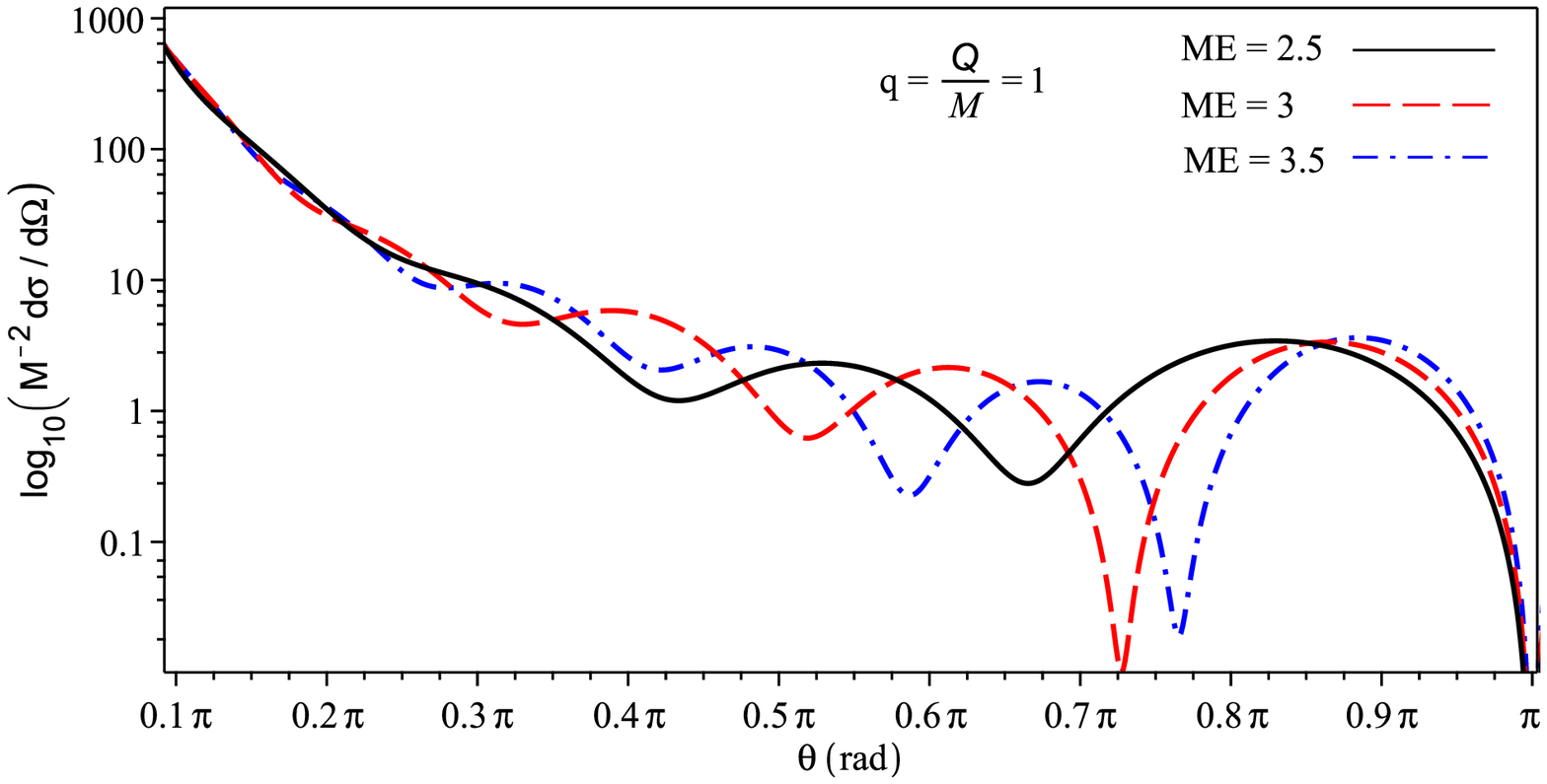}
\figcaption{\label{fig3}(color online). Reissner-Nordstr\"om scattering intensity for $ME=2.5,\ 3,\ 3.5$ in the case of a black hole with no charge $q=0$, with charge $q=0.5$ and the extremal case with charge $q=1$. Increasing the frequency has as an effect the narrowing of the glory width. At the same time the oscillations in the scattering intensity become more frequent. }
\end{center}

Fig. \ref{fig4} shows the behaviour of the massless fermion differential scattering cross section at low frequency for a typical Reissner-Nordstr\"om black hole ($q=0.5$). In Fig. \ref{fig5} the extremal Reissner-Nordstr\"om case ($q=1$) is studied for a large variance of $ME$. We notice the absence of oscillations at very low frequency ($ME=0.1$) in the differential scattering cross section. However, as the value of $ME$ is increased the spiral scattering and eventually glory start to occur.

\begin{center}
\includegraphics[scale=0.45]{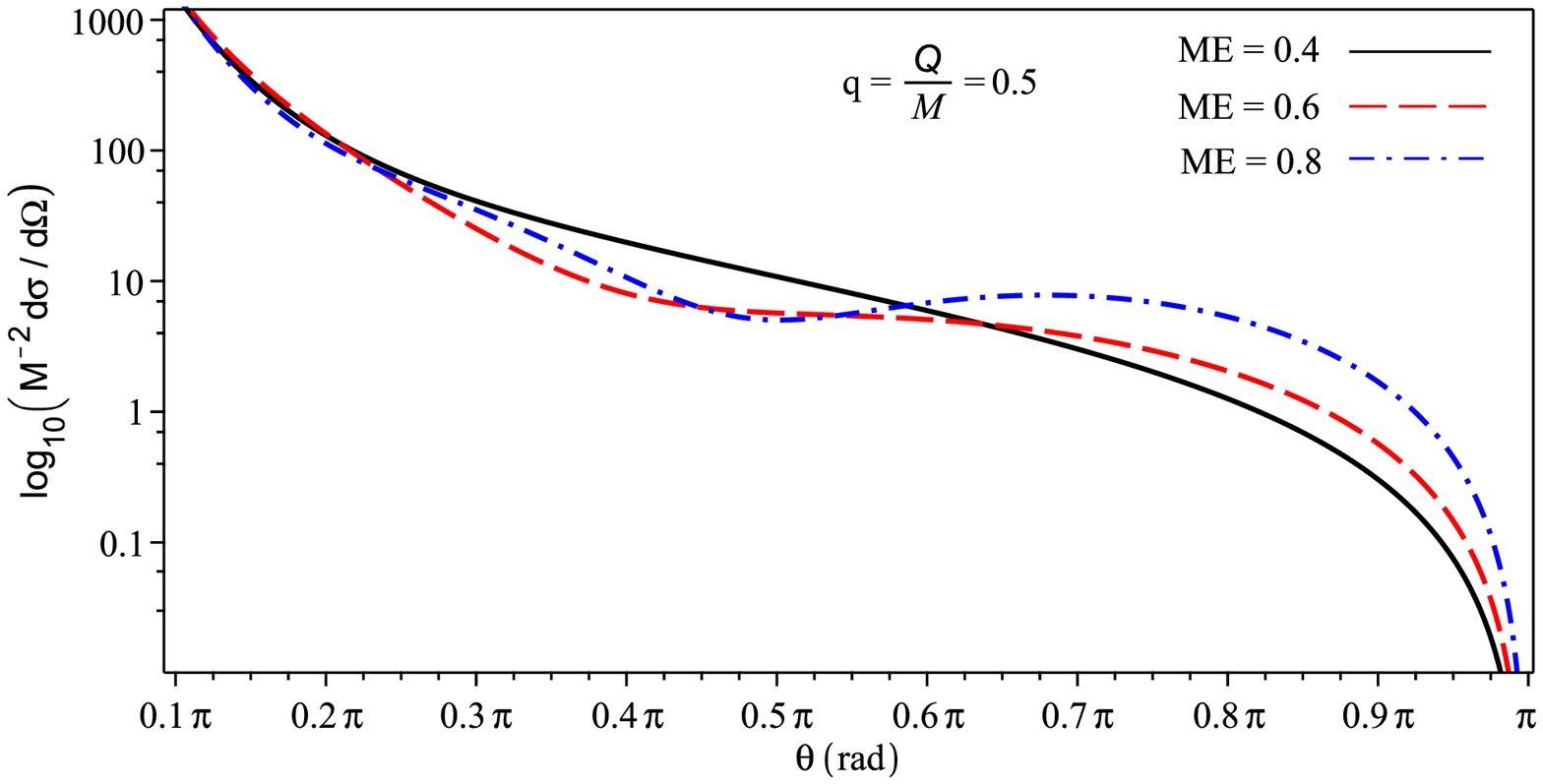}
\figcaption{\label{fig4}(color online). Differential scattering cross section for massless fermions at low frequencies ($ME=0.4,\, 0.6,\, 0.8$) for a typical Reissner-Nordstr\"om black hole with charge $q=0.5$. }
\end{center}

\begin{center}
\includegraphics[scale=0.45]{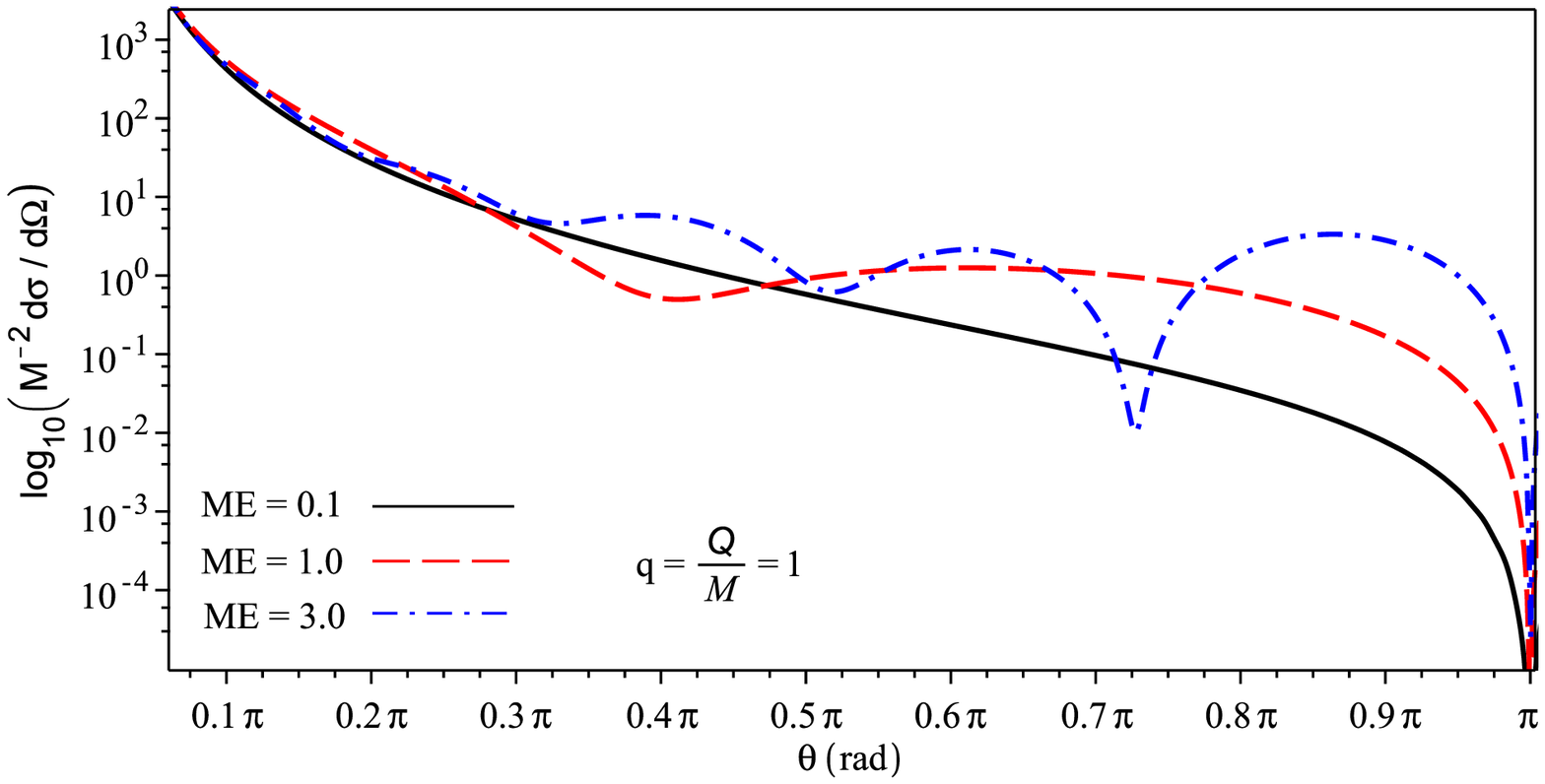}
\figcaption{\label{fig5}(color online). Scattering cross section for extremal Reissner-Nordstr\"om black hole ($q=1$) for $ME=0.1,\, 1,\, 3$. Increasing the value of $ME$, spiral scattering and glory start to occur in the scattering intensity.}
\end{center}

Exploring the parameter space $(q,\, ME)$ we have found situations when two different sets of parameters present similar scattering patterns (see Fig. \ref{fig6}). In some cases (like in Fig. \ref{fig6}B) the scattering patterns are almost the same. For example the difference between the values in the scattering intensity of $(0.6,\, 1.1)$ and $(1.0,\, 0.2)$ curves is less than 17\%. As a consequence one will need higher accuracy in the observed data in order to distinguish between the two data sets.

\begin{center}
\includegraphics[scale=0.45]{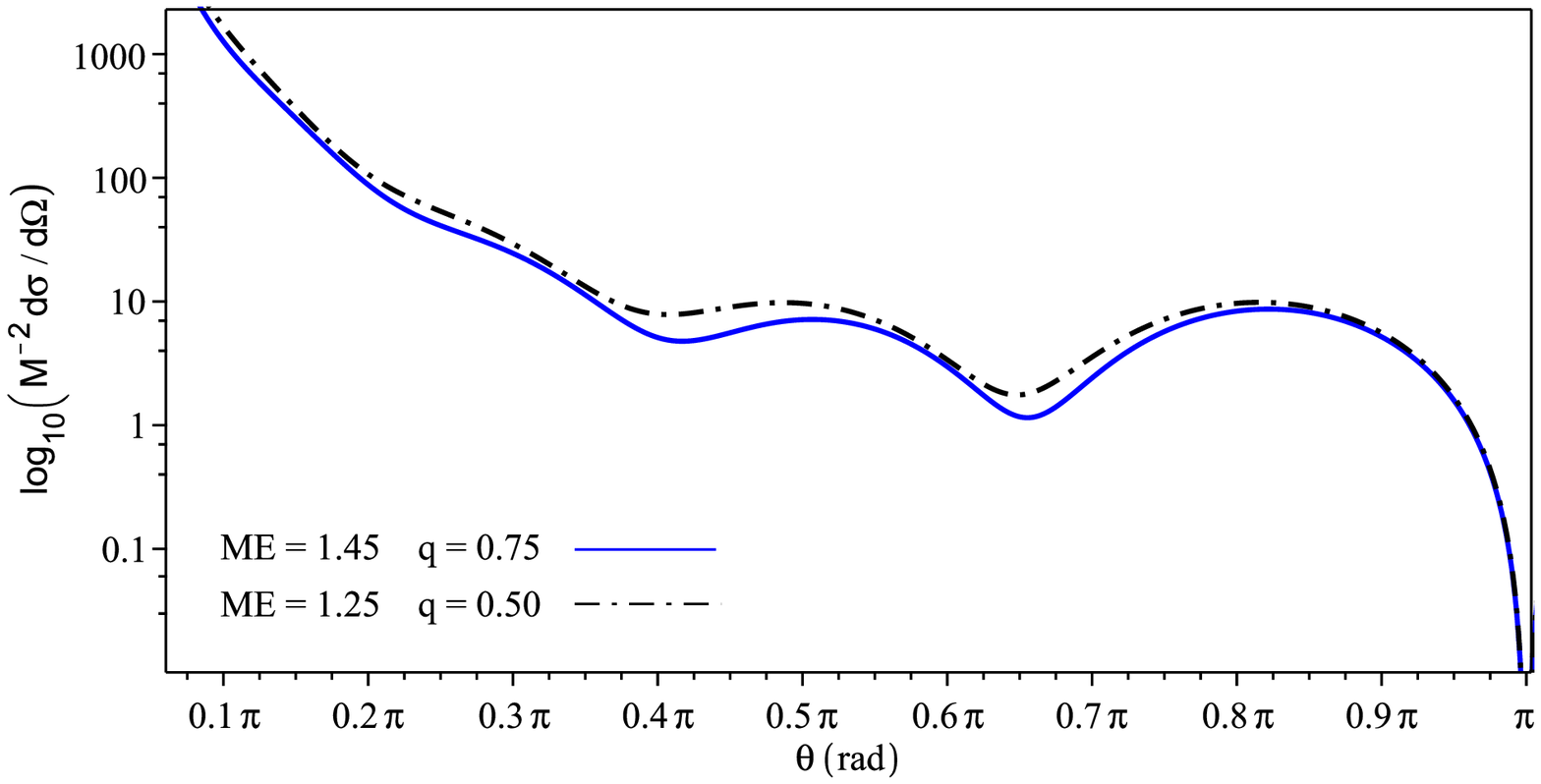}
\includegraphics[scale=0.45]{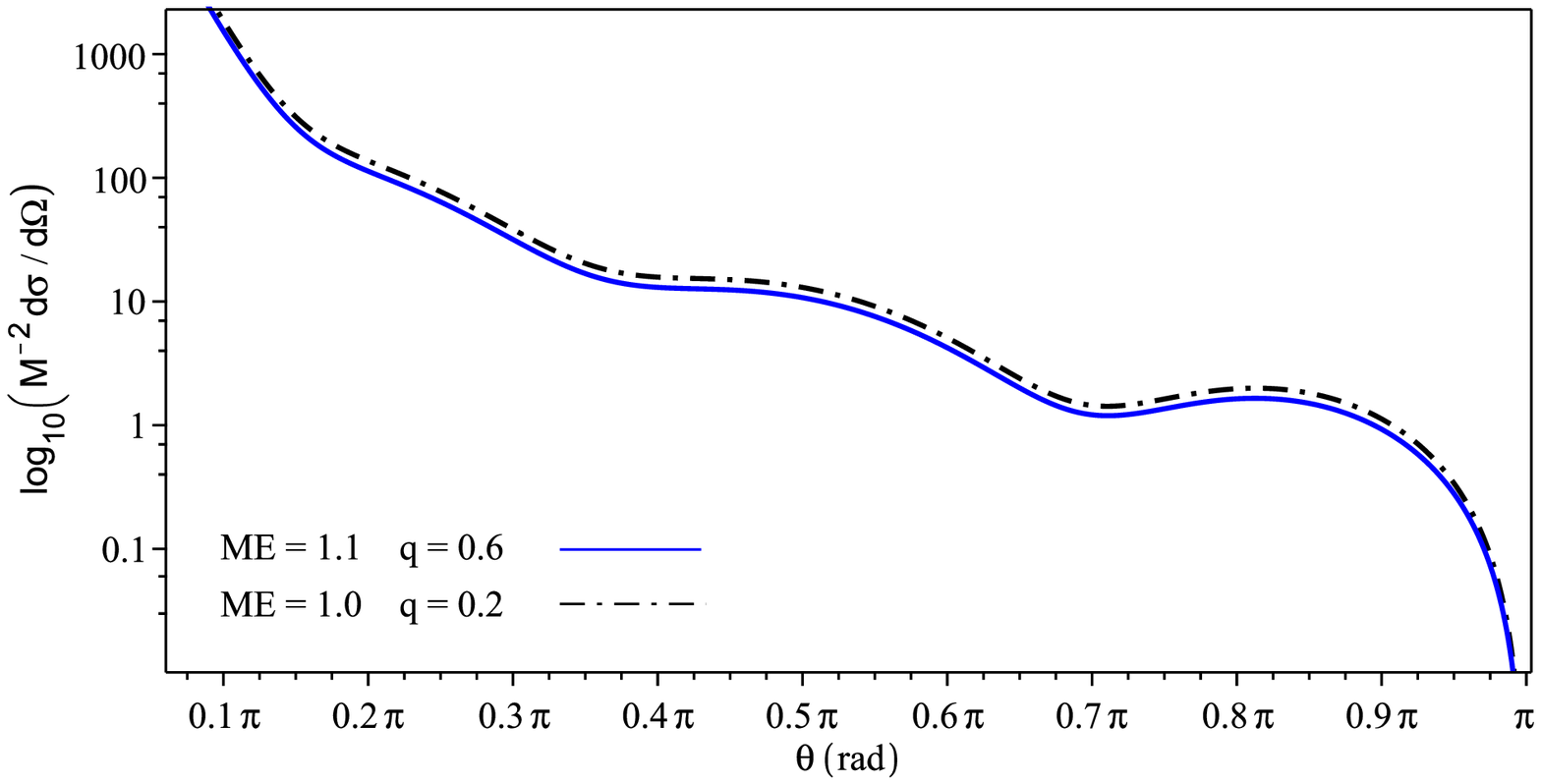}
\figcaption{\label{fig6}(color online). The dependence of the scattering pattern with the black hole charge and with the frequency. One can observe the similarities between the scattering patterns for certain values in the parameter space $(q,\, ME)$. }
\end{center}

\section{Conclusions and final remarks}\label{section6}

In this paper we have studied the scattering of massless fermions by Schwarzschild and charged Reissner-Nordstr\"om black holes. We have showed that glory and spiral scattering phenomena could occur for both types of black holes analysed, similar to what happens in the case of massive fermion scattering by black holes \cite{sporea1,sporea2,dolan}. As can be seen from Fig. \ref{fig1}-\ref{fig5} the scattering of massless fermions has always a minima in the backward direction (opposed to the massless scalar case \cite{crispino}). However, if the fermion becomes massive (see Fig. \ref{fig1}), then this minima starts to increase and will become eventually a maxima \cite{sporea1}.

The dependence of the scattering on the charge-to-mass ratio $q=Q/M$ was analysed for typical values including the extreme case $q=1$. As showed in Fig. \ref{fig2} for a fixed value of $ME$ the glory width gets larger as the value of $q$ is increased. One the other hand, keeping fixed the value of $q$ and varying the frequency $ME$ we observe an increase in the number of oscillations present in the scattering intensity, as showed in Fig. \ref{fig3}.

As already mentioned, at a fixed frequency, the glory pick is wider in the case of Reissner-Nordstr\"om black hole ($q\neq0$) compared with a Schwarzschild black hole ($q=0$). As a consequence the glory phenomena will be more easily to observe it astronomically for a Schwarzschild black hole than for a charged Reissner-Nordstr\"om one. Moreover the glory for extremal Reissner-Nordstr\"om case is the hardest one for astronomy observation.

We have used the parameter $ME$ to label our figures. Restoring the units we can make the following dimensionless quantity
\begin{equation}
\epsilon=\frac{GME}{\hbar c^3}=\frac{\pi r_S}{v\lambda_C}
\end{equation}
where $r_S=2MG$ is the Schwarzschild gravitational radius, $\lambda_C=h/p$ is the associated Compton wavelength of the particle and $v=p/E$ it's speed ($v=1$ for massless fermions). One can interpret $\epsilon$ as a measure of the gravitational coupling. The results obtained in the previous sections show that glory and spiral scattering of massless fermion by black holes are significant when the gravitational coupling is of order of $\pi$. This implies that we must have $r_S\sim\lambda_C$. Thus we can conclude that diffraction patterns of massless fermions (like the glory and spiral scattering) by black holes are significant if the condition $r_S\sim\lambda_C$ is fulfilled.

Neutrinos have the smallest mass among the fermions known experimentally today. The current upper bound limit on the sum of the three known neutrinos is of $\sum m_\nu < 0.183\,eV$ \cite{neutrinomass}. If we assume the mass of the electron neutrino to be of $m_{\nu_e}\sim0.01\,eV$, then the condition $r_S\sim\lambda_C$ implies a black hole mass of $M\sim10^{22}$ kg, which is much smaller than the mass of an astrophysical stelar black hole $M_{BH}\sim 10^{31}$ kg. This means that neutrino glory and spiral scattering can be observed only for scattering by small black holes. Such types of primordial black holes could have been created in the very early universe. Another possible scenario for the existence of such small black holes is in the context of theories with large extra-dimensions \cite{ADD}. In these circumstances the possibilities of observing and detecting diffraction patterns for massive fermion scattering by black holes are currently unavailable. However, in the case of existence of truly massless fermions (yet to be detected) we are no longer bound by the mass of the fermion (that as we saw constrains also the possible mass of the black hole), which means that glory and spiral scattering can in principle be observed for scattering of massless fermions (having appropriate energies) by real astrophysical black holes.

\section*{Acknowledgements}

I would like to thank Prof. I.I. Cot\u aescu for very useful discussions related to this subject, that helped to improve the manuscript and for suggesting to include Appendix A. I am also grateful to C. Crucean for discussions and for reading the manuscript. Last but not least, I would like to express my gratitude to the anonymous Referee for his/her comments and for suggesting to search for situations with different charge and frequency that present similar scattering patterns.

This work was supported by a grant of the Ministry of   National   Education   and   Scientific   Research,   RDI   Programme   for   Space Technology and Advanced Research - STAR, project number 181/20.07.2017.

\subsection*{Appendix A}\label{AppA}

\begin{small}

\noindent{\bf Neutrino limit to the Dirac field}

The aim of this Appendix is to show that the scattering of the Standard Model neutrino (which is a left-handed massless Dirac fermion) by black holes is contained in our results presented here regarding the scattering of massless Dirac fermions.

The following combination of the $(f^+_\kappa, f^-_\kappa)$ radial wave functions
\begin{subequations}
\renewcommand{\theequation}{A\arabic{equation}}
\begin{equation}
\begin{split}
& f^L_\kappa=\frac{1}{\sqrt{2}}\left( f^+_\kappa - if^-_{-\kappa} \right) \\
& f^R_\kappa=\frac{1}{\sqrt{2}}\left( f^+_\kappa + if^-_{-\kappa} \right)
\end{split}
\end{equation}
correspond to the radial wave functions for a left-handed fermion, respectively a right-handed one.

From eqs. (\ref{ec11}), (\ref{d3}) and $(f^+, f^-)^T=M(\hat f^+, \hat f^-) $ one gets
\begin{equation}
\begin{split}
f^+_\kappa &=i\sqrt{\varepsilon}\left(\hat f^-_\kappa - \hat f^+_\kappa \right) \\
&=i\sqrt{\varepsilon}\frac{1}{x}\left[\frac{s-i\varepsilon}{\kappa}C_1^{(\kappa)}M_{\rho_+,s}(2i\varepsilon x^2)- C_1^{(\kappa)}M_{\rho_-,s}(2i\varepsilon x^2) \right]
\end{split}
\end{equation}
\begin{equation}
\begin{split}
f^-_{-\kappa} &=\sqrt{\varepsilon}\left(\hat f^-_{-\kappa} + \hat f^+_{-\kappa} \right) \\
&=\sqrt{\varepsilon}\frac{1}{x}\left[\frac{s-i\varepsilon}{-\kappa}C_1^{(-\kappa)}M_{\rho_+,s}(2i\varepsilon x^2)+ C_1^{(-\kappa)}M_{\rho_-,s}(2i\varepsilon x^2) \right]
\end{split}
\end{equation}
where we used the condition $C_2=0$ as already mentioned in the main text for obtaining the phase shifts (\ref{final}).

Now by imposing the condition $f^R_\kappa=0$ and having in mind eq. (\ref{d4}) together with the fact that $M_{\rho_+,s}$ and $M_{\rho_-,s}$ are linearly independent, we obtain in the end the following relation
\begin{equation}\label{A}
C_1^{(\kappa)}=C_1^{(-\kappa)}
\end{equation}

Calculating now $f^L_\kappa$ using eq. (\ref{A}) we obtain in the end
\begin{equation}
f^L_\kappa=\sqrt{2}f^+_\kappa
\end{equation}

The last equation tells us that by applying the partial wave analysis to the left-handed neutrino $f^L_\kappa$ we will get the same phase shifts (\ref{final}) as obtained by applying the partial wave analysis to the fermion spinor $f^+_\kappa$.

\end{subequations}

\end{small}
\end{multicols}

\vspace{-1mm}
\centerline{\rule{80mm}{0.1pt}}
\vspace{2mm}

\begin{multicols}{2}

\end{multicols}

\clearpage
\end{CJK*}
\end{document}